\documentclass[
 reprint,
frontmatterverbose, 
nofootinbib,
 amsmath,amssymb,
 aps,
prd,
floatfix,
]{revtex4-1}

\usepackage{graphicx}

\usepackage{dcolumn}
\usepackage{bm}
\usepackage{pifont}
\newcommand{\cmark}{\ding{51}}%
\newcommand{\xmark}{\ding{55}}%

\usepackage{graphics}

\usepackage{hyperref}
\usepackage{xcolor,colortbl}
\definecolor{Gray}{gray}{0.75}

\begin{document}

\preprint{}

\title{\texorpdfstring{Getting chirality right: single scalar leptoquark solution/s to the $(g-2)_{e,\mu}$ puzzle}{Getting chirality right: single scalar leptoquark solution/s to the (g-2)e/mu puzzle}}

\author{Innes Bigaran}
\email{innes.bigaran@unimelb.edu.au}
 \affiliation{ARC Centre of Excellence for Particle Physics at the Terascale, School of Physics, The University of Melbourne, Victoria 3010, Australia}
\author{Raymond R. Volkas}
 \email{raymondv@unimelb.edu.au}
 \affiliation{ARC Centre of Excellence for Particle Physics at the Terascale, School of Physics, The University of Melbourne, Victoria 3010, Australia}

\date{\today}
\begin{abstract}
We identify the two scalar leptoquarks capable of generating sign-dependent contributions to leptonic magnetic moments, $R_2\sim (\mathbf{3}, \mathbf{2}, 7/6)$ and $S_1\sim (\mathbf{3}, \mathbf{1}, -1/3)$, as favoured by current measurements. We consider the case in which the electron and muon sectors are decoupled, and real-valued Yukawa couplings are specified using an up-type quark mass-diagonal basis. Contributions to $\Delta a_e$ arise from charm-containing loops and $\Delta a_\mu$ from top-containing loops -- hence avoiding dangerous LFV constraints, particularly from $\mu \to e \gamma$. The strongest constraints on these models arise from contributions to the Z leptonic decay widths, high-$p_T$ leptonic tails at the LHC, and from (semi)leptonic kaon decays.  To be a comprehensive solution to the $(g-2)_{e/\mu}$ puzzle we find that the mass of either leptoquark must be $\lesssim 65$ TeV. This analysis can be embedded within broader flavour anomaly studies, including those of hierarchical leptoquark coupling structures. It can also be straightforwardly adapted to accommodate future measurements of leptonic magnetic moments, such as those expected from the Muon $g-2$ collaboration in the near future. 
\end{abstract}

\maketitle

\section{\label{sec:Intro}Introduction}

The remarkable agreement between measurements and predictions of the muon and electron magnetic dipole moments has long been testament to the success of quantum field theory. Precise measurements of the deviation of this observable from the classical, tree-level value, $g_\ell=2$, give a sensitive probe of higher-order effects -- within the Standard Model~(SM) and beyond. SM corrections are precisely known, and therefore these specify the quantity $a_{\ell}^\text{SM}$, where
\begin{equation}
a_{\ell} \equiv {\small\frac{1}{2}}{(g-2)_\ell}.
\end{equation}
This makes anomalies in $(g-2)_\ell$, particularly if these differ between lepton flavour, a very strong indication of new physics~(NP) effects at loop-level\cite{Giudice:2012ms, Campanario:2019mjh}.

For the muon, there is persistent deviation between the SM prediction and the measured value~\cite{Blum:2013xva,Chapelain:2017syu},
\begin{equation}\Delta a_\mu= a_\mu^{\text{exp}} - a_\mu^{\text{SM}},
\end{equation}
corresponding to a  $3.6\sigma$ anomaly\footnote{The uncertainty values refer to the experimental and theoretical prediction uncertainties, respectively. See also the `{note added}' at the end of the conclusion.}:
\begin{equation}
\Delta a_\mu= (286\pm63\pm 43) \times 10^{-11}. \label{g-2mu}
\end{equation}
 Similarly, recent experimental results have indicated a deviation for the electron magnetic moment, of $2.5~\sigma$ significance~\cite{articleParker}:
\begin{equation}
\Delta a_e= -(0.88\pm0.36) \times 10^{-12}.\label{g-2e}
\end{equation}
{\small
  \begin{table}[t!]
\centering
 \renewcommand*{\arraystretch}{1.4}
\begin{tabular}{ |c|c|c|c| }
 \hline
 Symbol & $SU(3)_C \otimes SU(2)_L \otimes U(1)_Y$ & $(g-2)_{\ell}$ at 1L & $|F|$\\
 \hline
$\tilde{S}_1$   & $ (\mathbf{3}, \mathbf{1}, -4/3)$  & \xmark & 2\\
 \rowcolor{Gray}
$S_1$   & $ (\mathbf{3}, \mathbf{1}, -1/3)$  & \cmark  & 2\\
$S_3$&   $ (\mathbf{3}, \mathbf{3}, -1/3)$ & \xmark & 2\\
$\overline{S}_1$   & $ (\mathbf{3}, \mathbf{1}, 2/3)$ & \xmark & 2\\
\rowcolor{Gray}
${R}_2$&   $ (\mathbf{3}, \mathbf{2}, 7/6)$ & \cmark & 0\\
$\tilde{R}_2$&   $ (\mathbf{3}, \mathbf{2}, 1/6)$ & \xmark & 0\\
 \hline
\end{tabular}
\caption{Scalar LQs and their transformation properties, under the hypercharge convention $Q=I_3+Y$. The second-last column indicates whether the model is able to generate one-loop~(1L) corrections to the muon and electron magnetic moments with opposite sign -- $i.e.$ the LQ has mixed-chiral couplings.}
 \end{table}}
It is important to note that the sensitivity of $a_\ell$ to NP at energy scale $\Lambda$ scales generally as $m_\ell^m/\Lambda^n$, for some integer $n,m$. This indicates that a heavier lepton generally provides a more sensitive probe of NP. However, due to the short lifetime of the tau, a precise measurement of its magnetic moment (and, consequentially, any deviation from the SM) is beyond the reach of current experiments. The important issue to be addressed in this paper is that the discrepancies, $\Delta a_\mu$ and $\Delta a_e$, are of \emph{opposite sign}, which is a difficulty to be overcome when searching for a common explanation\footnote{See, for example, references \cite{Liu:2018xkx,Crivellin:2018qmi, Abdullah:2019ofw,CarcamoHernandez:2019ydc, Crivellin:2019mvj,Endo:2019bcj,Badziak:2019gaf,Hiller:2019mou,Cornella:2019uxs, Haba2020:2002.10230v1,Dorsner:2020aaz} for alternative methods to explain these anomalies in a single model.}.

 The leading candidates to explain these deviations involve flavour-dependent, loop-level, NP effects. It has long been established that exotic scalar-only extensions to the SM are capable of generating sizeable corrections to $(g-2)_\ell$. Of particular interest are scalar leptoquark~(LQ) models, which have proven to be useful for reconciling other well-known flavour-dependent anomalies (\emph{e.g.} see references~\cite{Dorsner:2016wpm,Becirevic:2016oho, Popov:2016fzr, Buttazzo:2017ixm, Angelescu:2018tyl}), provide a portal to generating radiative neutrino mass~\cite{Cai:2017jrq}, and are embedded in a number of theories of unification (\emph{e.g.}\ see reference~\cite{PhysRevD.8.1240}).

 \subsection{\label{sec:IntroA} Chirality of scalar LQ models}

To begin characterising scalar LQ models, we first need to introduce some terminology. Motivated by the introduction of direct lepton-quark couplings, rather than separately considering lepton~(L) or baryon~(B) number conservation, these are absorbed into the definition of a new conserved quantity~\cite{articlePSFN} \emph{fermion} number, $F$;
\begin{align}
     F=3B+L.
 \end{align}
$F$ is well-defined for each of the finite number of LQ models, and characterises the types of interactions mediated. The $|F|=2$ LQs couple to multiplets of the form $\ell q$, and $|F|=0$ couple to $\overline{\ell} q$ \cite{Hewett:1997ce} . Table I gives an overview of the scalar LQs and their gauge-group transformations, adopting the symbol notation from reference~\cite{Dorsner:2016wpm}.

The important characteristic of models that can generate contributions to $(g-2)_{(e /\mu)}$ that are consistent with experiment is the \emph{chirality} of their Yukawa couplings. To generate one-loop corrections whose sign can vary between lepton flavours, the LQ must have \emph{mixed-chiral} couplings, \emph{i.e.}\ both left- and right-handed couplings to charged leptons are present. To see why this is, we begin with a brief overview of the established calculations for $a_\ell$ corrections from scalar LQ states.
 \begin{figure}[t]
\centering
  \includegraphics{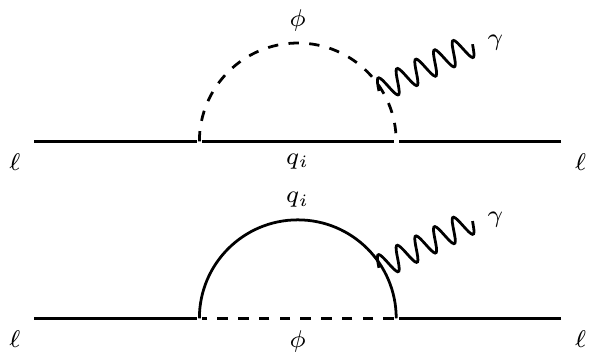}   \caption{ Dominant contributions to the lepton magnetic moment from scalar LQs. Arrows indicating fermion flow are omitted as there are multiple valid assignments possible for these topologies, each of which will be considered in calculations.}  \label{feyndi}
    \end{figure}

\subsection{\label{sec:IntroB}\texorpdfstring{Scalar LQs for $(g-2)_\ell$}{Scalar LQs for (g-2)l}}
In this section, we follow the calculation procedure for $\ell \to \ell' \gamma$ from reference \cite{Dorsner:2016wpm}, but adapt it specifically for $(g-2)_\ell$. The generic effective Lagrangian corresponding to contributions to $a_\ell$ is given by:
\begin{equation}
\begin{aligned}
    \mathcal{L}_{a_\ell}&=  e\overline{\ell} \left( \gamma_\mu A^\mu + \frac{a_\ell}{4m_\ell} \sigma_{\mu\nu}F^{\mu\nu} \right)\ell,\\
    &\subset  e\overline{\ell}  \gamma_\mu A^\mu  \ell +\frac{1}{2} i e\overline{\ell}  \sigma_{\mu\nu}F^{\mu\nu}\left( \sigma_L^\ell P_L+ \sigma_R^\ell P_R\right)\ell. \label{generical}
  \end{aligned}
\end{equation}
where $F^{\mu\nu}= \partial_\mu A_\nu - \partial_\nu A_\mu$, and $\sigma_{L/R}^\ell$ parameterise the effective left- and right-chiral interactions. Equation \eqref{generical} reveals, via coefficient matching, that
\begin{equation}
\Delta a_\ell= i m_\ell (\sigma_L^\ell+\sigma_R^\ell).\label{core1}
\end{equation}
 For the purpose of this discussion, for general scalar LQ $(\phi)$ models, the couplings to charged-leptons~($\ell$) and quarks~($q$) can be expressed as:
\begin{equation}
    \mathcal{L}_\ell = \overline{\ell^{(c)}}\; \left[y^{R}P_R + y^{L}P_L\right]\;q\;\phi^\dagger+h.c. \label{genericLQ}
\end{equation}
where $\ell^{(c)}=\ell$ for $|F|=0$ and $\ell^{(c)}=\ell^c$ for
$|F|=2$. 

We consider the two leading-order topologies for these corrections illustrated in Figure~\ref{feyndi}. Their contributions to $\sigma_{L/R}^\ell$ are well established in the literature~\cite{Dorsner:2016wpm}. For $F=0$ scalar LQs:

\begin{equation}
\begin{tiny}
    \sigma^{\ell}_L 
    = \frac{iN_c}{16\pi^2 m_{\phi}^2}
    \sum_{q} \left[m_\ell\left(|y^R_{\ell q}|^2+|y^L_{\ell q}|^2\right)\kappa+y^{R}_{\ell q} y^{L*}_{\ell q} m_q \kappa'\right], \label{sigmaL}
\end{tiny}
\end{equation}

\noindent and $\sigma^{\ell}_R=- [\sigma^{\ell}_L]^*$. In equation \eqref{sigmaL},
\begin{equation}
\begin{aligned}
&\kappa(x_q)= Q_{\phi}f_S(x_q)-f_F(x_q),\;\\  &\kappa'(x_q)= Q_{\phi}g_S(x_q)-g_F(x_q),\;\label{kappas}
\end{aligned}
\end{equation}
where $x_q=m_q^2/m_\phi^2$.  These contributions are proportional to the number of colours, $N_c=3$, and are summed over quark flavours $q$ running in the loop. The electric charge of the field $\phi$ is given by $Q_{\phi}$, and the loop functions in \eqref{kappas} are~\cite{Lavoura:2003xp, Dorsner:2016wpm}: 
{\small
\begin{equation}
\begin{aligned}
f_S(x)&= \frac{x+1}{4(x-1)^2}- \frac{x\log(x)}{2(x-1)^3},\\
f_F(x)&= \frac{x^2-5x-2}{12(x-1)^3}+ \frac{x\log(x)}{2(x-1)^4},\\
g_S(x)&=\frac{1}{x-1} -\frac{\log(x)}{(x-1)^2},\\
g_F(x)&= \frac{x-3}{2(x-1)^2}+\frac{\log(x)}{(x-1)^3}.
\end{aligned}\label{loopfun}
\end{equation}
}
 Therefore, we conclude via equation \eqref{core1} that
{\small
\begin{equation}
\begin{aligned}
    \Delta a_\ell = -\frac{3m_\ell}{8\pi^2 m_{\phi}^2} \sum_{q} \Big[m_\ell(&|y^R_\ell|^2+|y^L_\ell|^2)\;\kappa(x_q)\\
    &+m_q \text{Re}(y^{L*}_\ell y^{R}_\ell)\;\kappa'(x_q) \Big] .\label{deltaa}
    \end{aligned}
\end{equation}}
 For a scalar LQ with \emph{maximally chiral} Yukawa couplings, the second term will not be present and contributions from each propagator will be of definite relative sign. However, for mixed-chiral scalar LQs, we can access terms proportional to $\kappa'$, allowing us to vary the sign of the NP contribution. As these contributions scale proportional to $m_q$, we expect the dominant contributions to be those with the third-generation quarks entering the loop. 
 
 Taking $m_\ell \ll m_{q}$, the like-handed terms are subdominant to the mixed-handed contributions -- exactly the terms required for generating contributions with relative-sign. This leaves the following:
\begin{equation}
\begin{aligned}
    \Delta a_\ell \sim -\frac{3m_\ell}{8\pi^2 m_{\phi}^2} \sum_{\{q:\; m_\ell \ll m_q\} }
    m_q \text{Re}(y^{L*}_{\ell, q} y^{R}_{\ell,q})  \;\kappa'(x_q).\label{alfinal}
    \end{aligned}
\end{equation}
For $|F|=2$ LQs, the above applies but with $Q_\phi \mapsto -Q_\phi$, $y^R_\ell \mapsto y^{L*}_\ell$ and $y^L_\ell \mapsto y^{R*}_\ell$.

For mixed-chiral scalar LQ models, this provides a portal to flavour-dependent sign allocation for the correction. We have the clear prospect of meeting current experimental measurements with simple, single-scalar field extensions\footnote{We consider only a single-scalar solution, whereas with multiple scalar LQ models the idea of using LQ mixing to generate (at least) $(g-2)_\mu$ was explored in reference \cite{Dorsner:2019itg}.}, such as those identified in Table I. Please note that we simply require the mixed-chiral term to \emph{dominate}, but that the like-chiral terms will also be included in later calculations (see Section III).

\subsection{\label{sec:IntroC}\texorpdfstring{Is there a no-go theorem for single scalar LQs to generate $(g-2)_{e/\mu}$?}{Scalar LQs for (g-2)l}}

As the mixed-chiral contributions in equation~\eqref{alfinal} are proportional to the mass of the quark in the loop, we may initially, naively, restrict LQ couplings to represent a \emph{top-philic} coupling texture -- maximising this $m_q$ enhancement. However, we will begin here by discussing why such a model is severely disfavoured.

In reference \cite{Crivellin:2018qmi}, the authors show that\footnote{We thank an anonymous referee for drawing our attention to this reference and associated proof.}, for a single-field extension of the SM, with the muon and electron sector not decoupled, the anomalies in $\Delta a_{\mu/e}$ are incompatible with the rare decay $\mu \to e \gamma$. For the scalar leptoquarks discussed in Section \ref{sec:IntroB}, from Figure~\ref{feyndi} if the quark ($q_i$) coupling to the leptoquark ($\phi$) is identical for both charged-leptons, then the $\mu\to e \gamma$ transition can be obtained by combining a subset of the leptoquark vertices involved in generating $\Delta a_{\mu/e}$. As a result, the following expression holds:
\begin{equation}
    \text{Br}[\mu \to e \gamma]= \frac{e^2 }{64\pi \Gamma_\mu} \frac{m_\mu^2}{m_e}|\Delta a_\mu \Delta a_e|\sim 9 \times 10^{-5} \label{crev}
\end{equation}
where $\Gamma_\mu$ is the full decay width of the muon. This algebraic expression is consistent with that quoted in \cite{Crivellin:2018qmi}, within the justifiable limit that $m_e~\ll~m_\mu$. The numerical value is a result of inputting the results from \eqref{g-2mu} and \eqref{g-2e}, which is eight orders of magnitude above the current experimental bound on this process from the MEG collaboration~\cite{TheMEG:2016wtm}:
\begin{equation}
     \text{Br}[\mu \to e \gamma]^\text{MEG}< 4.2\times 10^{-13}.
\end{equation}

For this reason, the single-leptoquark solution to this process seems to be heavily disfavoured. However, the validity of equation \eqref{crev} relies on the assumption that the quark in the loop of Figure \ref{feyndi} is the same for both the muon and electron $\Delta a_\ell$ contributions. This no-go result could be avoided by relaxing this assumption. 

Without a like-quark coupling between the leptoquark and both charged-lepton generations, we can avoid the constraint from $\text{Br}[\mu \to e \gamma]$ by restricting any contributions to be a higher-order process. The same philosphy applies to other lepton-flavour violating processes such as $Z\to e \mu$, $K_L\to e \mu$ and muon-electron conversion in nuclei. The relative success of such a restriction is dependent on the flavour ansatz for Yukawa couplings, as will be further discussed in Section~\ref{subsecII1}. Although this decoupling may reduce the dominance of the mixed-chiral term to $\Delta a_\ell$, it remains to be seen whether such a model could still be successful.

\section{\label{sec:Models}Models of interest}

As summarised in Table I, the $S_1$ and $R_2$ leptoquarks are able to induce opposite-sign $(g-2)_{e/\mu}$ contributions via the mixed-chiral contribution in equation \eqref{alfinal}. These LQs have also garnered recent attention in other flavour anomaly studies \cite{Bauer:2015knc,Crivellin:2017zlb, Cai:2017wry, Chauhan:2017ndd,Bigaran:2019bqv, Crivellin:2019dwb}. The relevant LQ couplings for each extension, represented here as $3\times 3$ Yukawa coupling matrices, are given by\footnote{We implement the interaction basis with neutrinos in their flavour eigenstates (\emph{i.e.} no PMNS rotation) and consider only terms in which these fields couple as leptoquarks. We impose conservation of baryon number  $U(1)_B$ to forbid diquark couplings.}:
\begin{equation}
 \begin{aligned}
 \mathcal{L}_{\text{int}}^{S_1} = & \left( \overline{L_L^c} \lambda_{LQ} Q_L +  \overline{e_R^c}\lambda_{eu} u_R\right) S_1^\dagger+ h.c.,
 \end{aligned}
 \end{equation}
  \begin{equation}
 \begin{aligned}
 \mathcal{L}_{\text{int}}^{R_2} &= \left( \overline{L_L}\lambda_{Lu} u_R +  \overline{e_R} \lambda_{eQ} Q_L \right)R_2^\dagger + h.c.
 \label{lagrange2}
 \end{aligned}
 \end{equation}
The doublet, $R_2$, can be expressed in terms of its electric charge-definite components: 
\begin{align}
    R_2 \sim \begin{pmatrix}R_2^{5/3} \\ R_2^{2/3} \end{pmatrix},
\end{align}
with charges as indicated by the superscripts. We assume negligible mass-splitting between the components of the multiplet, \emph{i.e}, $$m_{R_2} \approx m_{R_2^{5/3}}\approx m_{R_2^{2/3}},$$ so as to avoid constraints from electroweak oblique corrections~\cite{Dorsner:2016wpm}.

\subsection{Coupling framework}
 \subsubsection{`Up-type' vs. `down-type' mass-diagonal basis for Yukawa couplings}\label{subsecII1}
 
 When rotated into the flavour eigenbasis, we have two choices to redefine the couplings: either an `up-type' or `down-type' mass-diagonal basis. Practically, this involves a choice of which couplings we `fix' to a particular texture, and which we allow to be generated by CKM mixing. 
 
 The `up-type' mass-diagonal Yukawas are defined in accordance with the mappings
\begin{equation}
 \begin{aligned}
 \mathfrak{R}_e \lambda_{eu}\mathfrak{R}_u\mapsto  y^{Seu},\;\;\;\mathfrak{L}_e \lambda_{LQ}\mathfrak{L}_u\mapsto y^{SLQ}, \\
 \mathfrak{L}_e^\dagger \lambda_{Lu}\mathfrak{R}_u\mapsto y^{RLu},\;\;\;
\mathfrak{L}_e^\dagger\lambda_{eQ} \mathfrak{L}_u \mapsto y^{ReQ}. \label{mappings}
 \end{aligned}
 \end{equation}
 
\noindent Here, $\mathfrak{L}$ and $\mathfrak{R}$ represent the basis mapping between the gauge and flavour eigenstates, and $V = \mathfrak{L}_u^\dagger \mathfrak{L}_d$ is the standard CKM matrix\footnote{As neutrino masses are negligible for the phenomenology of this paper, we keep neutrinos in the flavour eigenbasis and set the PMNS matrix to the identity throughout all subsequent calculations.}. 
 
 By `mass-diagonal' we refer to the mathematical outcome of this choice. The non-physical left-handed rotation matrix ($\mathfrak{L}$) for either the `up-type' or `down-type' quarks is set to the identity, and we allow the CKM to be correspondent directly with the $\mathfrak{L}$ of the alternative quark-type.
 
In the `up-type' formalism, the relevant down-type quark couplings are related to those of the up-type quarks via the CKM matrix. This flavour ansatz allows us to explicitly forbid one-loop contributions to LFV processes by decoupling the muon and electron sectors.

The alternative basis choice would be to select the down-type quark couplings to fix, and allow associated up-type quark couplings to be generated via CKM mixing, thus defining the `down-type' mass diagonal basis. This formalism for tackling $(g-2)_{e/\mu}$, for which models with single scalar leptoquarks have been shown to be ruled-out by significant contribution to $\mu \to e \gamma$~\cite{Dorsner:2020aaz}.

Of course, the validity of a particular field content in NP models is basis independent, however we typically model-build around choices of specific non-zero coupling textures. This may lead to missing allowed parameter space if one were to consider only one of these two Yukawa coupling bases. 

 \subsubsection{Model framework under the `up-type' flavour ansatz}
 \label{subsecII2}

There are two independent coupling matrices for each model, and the interaction Lagrangians may be re-expressed as:

\begin{equation}
 \begin{aligned}
 \mathcal{L}^{S_1} \supset y^{SLQ}_{ij} \left[ \overline{e_{L,i}^c} u_{L,j} - V_{jk} \;\overline{\nu_{L,i}^c} d_{L,k} \right]S_1^\dagger\\ + y^{Seu}_{ij} \overline{e_{R,i}^c} u_{R,j} S_1^\dagger+ h.c., \label{lags}
 \end{aligned}
 \end{equation}
   \begin{equation}
 \begin{aligned}
 \mathcal{L}^{R_2} \supset &y^{RLu}_{ij} \left[ \overline{\nu_{L,i}} u_{R,j} R_2^{2/3, \dagger} -\; \overline{e_{L,i}} u_{R,j} R_2^{5/3, \dagger} \right]\\
 &+y^{ReQ}_{ij} \overline{e_{R,i}}\left[ u_{L,j}R_2^{5/3, \dagger} +V_{jk}   d_{L,k}R_2^{2/3, \dagger}\right]\\
 &+ h.c.\label{lagr}
 \end{aligned}
 \end{equation}
Recalling the discussion in Section I, we note that $R_2^{5/3}$, and $S_1$ both have left- and right-handed couplings to charged leptons and SM up-type quarks. It is these couplings which we have deemed most important for the $(g-2)_\ell$ anomalies, and so we will fix the up-type couplings to $\phi$, and allow down-type interactions to be generated only by virtue of their model-dependent relationships to these. We will discuss these couplings further in Section II-A.

The parameters $\kappa'$ for each model, as per \eqref{kappas}, are given by the following, assuming $m_q^2\ll m_\phi^2$:
{\small
\begin{equation}
\begin{aligned}
   \kappa'_{S_1}(m_q)&= \frac{7}{6}+ \frac{2}{3}\log\left(\frac{m_q^2 }{m_{S_1}^2}\right),\\
    \kappa'_{R_2}(m_q)&= -\frac{1}{6}- \frac{2}{3}\log\left(\frac{m_q^2}{m_{R_2}^2}\right).
    \end{aligned}
\end{equation}}

\noindent Their mixed-chiral $\Delta a_\ell$ contributions are therefore, via equation \eqref{alfinal}, given by:
\begin{small}
\begin{align}
 \Delta a_\ell^{S_1} \sim -\frac{m_\ell m_q}{4\pi^2 m_{S_1}^2}
     \left[\frac{7}{4}- 2\log\left(\frac{m_{S_1}}{m_q}\right)\right]\text{Re}(y^{L*}_{\ell q} y^{R}_{\ell q}), \label{ALe}
 \end{align} 
  \begin{align} 
 \Delta a_\ell^{R_2} \sim \frac{m_\ell m_q}{4\pi^2 m_{R_2}^2} \left[\frac{1}{4}-2\log\left(\frac{m_{R_2}}{m_q}\right)\right] \text{Re}(y^{L*}_{\ell q} y^{R}_{\ell q}),\label{ALm}
  \end{align} 
\end{small}

\noindent where for $S_1$, 
\begin{align}
    \mathbf{y}^R =\mathbf{y}^{Seu}\; \text{and}\; \mathbf{y}^L= \mathbf{y}^{SLQ},\label{coup1}
\end{align} and for $R_2$,
\begin{align}
\mathbf{y}^R=-\mathbf{y}^{RLu}\; \text{and}\;\mathbf{y}^L=\mathbf{y}^{ReQ}.\label{coup2}
\end{align}
For consistency between $|F|=0$ and $|F|=2$ models, and with equation \eqref{generical}, we have labelled the chirality to be that of the quark field in the associated interaction term. Furthermore, the strongest contributions in the $R_2$ model arise only from interactions of the $R^{5/3}$ component of the LQ doublet. The component $R^{2/3}$ does not have couplings of both chiralities to charged leptons.

 Input quark masses are the $\overline{\text{MS}}$ masses at the scale where each contribution is `integrated out', \emph{i.e} $m_c(m_c)$ and $m_t(m_t)$. The Yukawa couplings here are at the high scale ($m_\phi$), where we have neglected running of these NP contributions between this and lower energy scales. We estimate via numerical trial that this introduces $\sim \mathcal{O}(10\%)$ effect. For larger LQ masses, a more careful EFT analysis may be preferable, but within the scope of this work we have neglected such effects for calculation of $\Delta a_\ell$.

\subsection{Coupling textures}
\label{sec:ModelsA}

The Yukawa coupling values themselves are \emph{a priori} completely arbitrary. The mixed chiral terms shown in \eqref{ALe} and \eqref{ALm} are directly proportional to the mass of the quark in the loop in Figure \ref{feyndi} -- so an enhancement of this term is best achieved by LQ couplings to higher-generation quarks. Ideally, contributions to the muon and electron magnetic moments would both be enhanced proportional to $m_t$; however, Section I-C has shown that this is incompatible with constraints from Br$(\mu \to e \gamma)$. The next-best thing is to couple these charged leptons separately to either the top or the charm; in both models, it is only the up-type quarks that have both chiralities of the required coupling.

In this framework, there are two options for textures to generate $\Delta a_\ell \neq 0$, $\ell \in \{e, \mu \}$, labelled Texture 1 and Texture 2. We reiterate that the convention adopted here is such that the rows are labelled as charged-lepton, and columns as up-type quark, generations. Here, non-zero couplings are shaded in grey:
{\small
 \begin{equation}
{\text{Texture 1}} \sim \begin{pmatrix} 0 & 0 & \colorbox{Gray}{\color{Gray}'}  \\ 
                                     0 &\colorbox{Gray}{\color{Gray}'}  & 0 \\
                                     0 & 0 & 0 
                                                                          \end{pmatrix},\;\;                   
                                     \text{Texture 2} \sim \begin{pmatrix} 0 &  \colorbox{Gray}{\color{Gray}'}& 0 \\ 
                                     0 &0  & \colorbox{Gray}{\color{Gray}'} \\
                                    0 & 0 & 0
                                    \end{pmatrix}.       \label{yukawas}  \end{equation}
   }                                 
For the mixed-chiral interaction to be present, and thus enhanced,  both the left- and right-handed couplings need to have the same nonzero texture. 

\subsubsection{\texorpdfstring{Texture 1: the Charmphilic solution for $(g-2)_\mu$}{Texture 1: Charmphilic solution}}
\label{sec:ModelsA1}
A \emph{charmphilic} model for $\Delta a_\mu$  was explored for both $R_2$ and $S_1$ by Kowalska \emph{et al} ~\cite{Kowalska:2018ulj}. Notably, for generating $\Delta a_\mu$ via charm-muon coupling, the $R_2$ model is found to be entirely ruled-out by a combination of flavour constraints and searches in the LHC dimuon channel~\cite{Kowalska:2018ulj}. As a solution for $(g-2)_\mu$, the muon-charm coupling for $S_1$ was explored and found to have highly constrained but nonzero allowed parameter space. 

For $S_1$ under Texture 1, the most constraining process on the muon NP couplings is from the decay $K^+ \to \pi^+ \nu \nu$. Couplings between the strange-quark and neutrinos are unavoidably generated via CKM mixing, where $V_{cs} \sim \mathcal{O}(1)$. For non-zero left-handed couplings to the charm, there will be non-negligible contributions to this decay width. Following the derivation from reference~\cite{Kowalska:2018ulj}, the $2\sigma$ bound on Br($K^+ \to \pi^+ \nu \nu$)(Table 1) generates the following constraint:
\begin{align}
    |y^{SLQ}_{22}|< 5.26 \times 10^{-2} \frac{m_\phi}{\text{TeV}}.
\end{align}

Note that larger LQ masses allow a weakening of this bound when applied to the coupling $y^{SLQ}_{22}$. However, such mass and coupling combinations are found to be heavily constrained by LHC dimuon channel searches, via $pp\to \mu\mu (j)$. The $S_1$ model is found to viable only within two-sigma of the $\Delta a_\mu$ central value, by a combination of flavour constraints and searches in the LHC constraints~\cite{Kowalska:2018ulj}. 

The results of reference~\cite{Kowalska:2018ulj} motivate consideration of the alternative coupling structure, Texture 2, in this work. For both $R_2$ and $S_1$, we will explore a novel, minimal parameter space, and extend the single-leptoquark model to both electron and muon magnetic moments.

\subsubsection{\label{subsec:1.5} Texture 2: establishing the focus of this work}
\label{sec:ModelsA2}
As argued above, the most sensible approach hereon will be to implement Texture 2 in \eqref{yukawas}, \emph{i.e.}\ for the remainder of this work we will consider the following nonzero couplings for both models:
{\small
\begin{equation}
\mathbf{y}^{L} \sim \begin{pmatrix} 0 &  \colorbox{Gray}{\color{Gray}'}& 0 \\ 
                                     0 &0  & \colorbox{Gray}{\color{Gray}'} \\
                                    0 & 0 & 0
                                    \end{pmatrix},\;\;\;\;
\mathbf{y}^{R} \sim \begin{pmatrix} 0 &  \colorbox{Gray}{\color{Gray}'}& 0 \\ 
                                     0 &0  & \colorbox{Gray}{\color{Gray}'} \\
                                    0 & 0 & 0
                                    \end{pmatrix}.                                
                                \label{yukawas2}  \end{equation} }

\noindent Also, for the remainder of this work we will restrict our input couplings to real values in order to circumvent constraints from $CP$-violating observables\footnote{The parameter space could, of course, be extended to complex couplings, but to do so we would need to carefully consider (among other things) the contribution of the diagrams in Figure \ref{feyndi} to electric dipole moments~\cite{Dekens:2018bci}.}.

Assuming that the mixed-chiral terms of \eqref{ALe} and \eqref{ALm} dominate, we begin by deriving rough analytic bounds on the products of relevant couplings. The current experimental values for $\Delta a_{e}$~\eqref{g-2e} and $\Delta a_{\mu}$~\eqref{g-2mu} yield the following for $S_1$:
\begin{small}
\begin{align}
 y^{L}_{23} y^{R}_{23}= -(0.35\pm 0.09) \times \left(\frac{ m_{S_1}}{10\text{TeV}}\right)^2,
 \label{bound1}
 \end{align} 
  \begin{align} 
  y^{L}_{12} y^{R}_{12}=(0.46\pm 0.19) \times \left(\frac{ m_{S_1}}{10\text{TeV}}\right)^2,
  \label{bound2}
  \end{align} 
\end{small}

\noindent and for $R_2$:
\begin{small}
\begin{align}
 y^{L}_{23} y^{R}_{23}=(0.20\pm 0.05)\times \left(\frac{ m_{R_2}}{10\text{TeV}}\right)^2,
 \label{bound3}
 \end{align} 
  \begin{align} 
  y^{L}_{12} y^{R}_{12}= -(0.41\pm0.17)\times \left(\frac{ m_{R_2}}{10\text{TeV}}\right)^2.
  \label{bound4}
  \end{align} 
\end{small}

\noindent
Note that the loop functions in~\eqref{loopfun} depend explicitly on the leptoquark mass, and so the $m_\phi$ dependence cannot truly be completely factored-out. The above bounds rely on the assumption that evaluating~\eqref{loopfun} for $m_\phi=1$~TeV gives a reasonable numerical approximation to their behaviour in a viable mass range.

Although relations~\eqref{bound1}-\eqref{bound4} guide us to the relative sizes of these couplings, we will proceed with a more careful phenomenological analysis to grasp the full scope of these models.
 
 \subsection{Consideration of relevant constraints}
\label{sec:ModelsB}
Regardless of which of the two models we consider, the crossing-symmetry between the topologies in Figure \ref{feyndi} and contributions of the form $\ell \to \ell' (\gamma/Z)$, we expect the strong constraints to be from $Z \to \ell \ell'$ processes \cite{Djouadi:1989md, Popov:2016fzr,ColuccioLeskow:2016dox,Das:2016vkr,Arnan:2019olv}. As we have decoupled the leptonic sectors, the strongest constraints of this form will be from LFU processes, particularly from muon-top couplings \emph{i.e} $\ell=\ell'=\mu$. By virtue of the structure of equations \eqref{lags} and \eqref{lagr}, couplings to neutrinos generate contributions to the invisible width of the Z. These effective electroweak couplings provide the most stringent constraints on muon couplings to the top-quark.

The couplings between the electron and the charm-quark give a more rich phenomenology. They are most strongly constrained by dimension six contact interactions, which manifest in high-$p_T$ lepton tails at the LHC, short-distance effects in leptonic charged-current charmed meson decays, and related constraints from the kaon sector\cite{Fuentes-Martin:2020lea}. Contributions to mesonic decays are parameterised using a low-energy EFT, and as such the Wilson coefficients~(WCs) corresponding to SMEFT operators outlined in Table \ref{table:SMEFT} are run and matched to low-energy prior to the calculation of observables. This is achieved using the combination of \texttt{Wilson}~\cite{Aebischer:2018bkb} and \texttt{Flavio}~\cite{Straub:2018kue} packages.

\subsubsection{Effective electroweak couplings}
\label{sec:ModelsB1}
We follow the calculation of effective $Z$ couplings to charged-leptons $\ell$, and associated observables, from reference~\cite{Arnan:2019olv}. The couplings $g_{f_{L(R)}}$ are the effective left- and right-handed couplings of the Z boson to fermions, $f$. The effective Lagrangian for describing these interactions is given by:
\begin{equation}
    \mathcal{L}^Z_\text{eff} = \frac{g}{\cos(\theta_W)} \sum_{i,j} \overline{f_i} \gamma^\mu\left[g^{ij}_{f_L}P_L+g^{ij}_{f_R}P_R \right]f_j Z_\mu,
\end{equation}
where $\theta_W$ is the weak-mixing angle, and for $i=j$, we relabel the coupling $g^{ii}_{f_{L(R)}}\equiv g^{f_i}_{L(R)}$ for clarity. 

The strongest contributions to these corrections come from top-containing loops that have the same topology as those in Figure~\ref{feyndi}, but with a Z rather than photon line attached. The same couplings are relevant for these contributions as to $\Delta a_\mu$-- which motivates careful consideration of $Z\to\mu\mu$ in our numerical study. For this study, we enforce $2\sigma$ agreement for the effective left- and right-handed couplings about the central values ($g^\mu_{L/R}$) quoted in Table~\ref{table:const}. We do not directly consider the correlation between these two parameters.

{\small
\begin{table}[t]
 \begin{center}
  \def\arraystretch{1.6} 
    \begin{tabular}{|c|c|c|} 
    \hline
      \textsc{Process}  & \textsc{Observable}&\textsc{Limits} \\
     \hline \hline
       $Z \to \ell_i \ell_j$ &  
          $ \{g^\mu_{L},g^\mu_{R}\}$ &$\{-0.2689(11),+0.2323(13)\}$\\
        \hline 
       $Z \to \nu \nu$              &  $N_\nu$  & $2.9840(82)$ \\
       \hline 
    $D^{\pm}\to e \nu$ & Br & $< 8.8 \times 10^{-6}$ \\
    \hline
    $D^{\pm}_s\to e \nu$ & Br &  $< 8.3\times 10^{-5}$  \\
    \hline
    $K^+\to \pi^+ \nu \nu$ & Br & $(1.7\pm 1.1) \times 10^{-10}$   \\ 
    \hline 
      $K^0_L\to e^+ e^-$ & Br & $(9^{+6}_{-4}) \times 10^{-12}$   \\ 
    \hline
    $pp\to \ell\ell$ &$|y^{Seu}_{12}|$& $< 0.648~m_\phi/\text{TeV}$\\
    &$|y^{SLQ}_{12}|$& $< 0.537~m_\phi/\text{TeV}$\\
     \cite{Greljo:2017vvb}&$|y^{ReQ}_{12}|$& $<0.393~m_\phi/\text{TeV}$\\
    &$|y^{RLu}_{12}|$& $<0.524~m_\phi/\text{TeV}$\\
    &$|y^{ReQ}_{23}|$& $<0.904~m_\phi/\text{TeV}$\\
    \hline 
     $c\to d_k \overline{e}_i \nu_j$& $\epsilon_{V_L}^{111}$& $\in[-0.52,0.86]\times 10^{-2}$\\
     & $ \epsilon_{V_L}^{112}$& $ \in[-0.28,0.59]\times 10^{-2}$\\ \cite{Fuentes-Martin:2020lea}
     & $\{|\epsilon_{V_L}^{121}|, |\epsilon_{V_L}^{122} |\}$& $\{< 0.67, < 0.42\}\times 10^{-2}$ \\
     & $\{|\epsilon_{S_L}^{1 1 1}|, |\epsilon_{S_L}^{1 1 2}|\}$ & $\{< 0.72 , < 0.43 \} \times 10^{-2}$
     \\
     & $\{|\epsilon_{S_L}^{2 1 1}|, |\epsilon_{S_L}^{2 1 2} |\}$ & $\{<1.1 ,<0.68 \} \times 10^{-2}$\\
    & $\{|\epsilon_{T}^{1 1 1}|, |\epsilon_{T}^{1 1 2}|\}$ & $\{<4.3 ,<2.8 \} \times 10^{-3}$
     \\
     & $\{|\epsilon_{T}^{2 1 1}|, |\epsilon_{T}^{2 1 2}| \}$ & $\{<6.6 ,<4.0\} \times 10^{-3}$\\
        \hline
   \end{tabular}
  \end{center}
    \caption{Processes most constraining on this model. Values quoted without citation are from PDG~\cite{PDG}.  Constraints from $pp\to ee$ are derived from Table 1 of reference~\cite{Greljo:2017vvb}, in conjunction with the expressions from Table~\ref{table:SMEFT}. For $c\to d_k \overline{e}_i \nu_j$, $95\%$ confidence intervals for parameters~\eqref{epsilonsSMEFT} are quoted where accessible, otherwise upper-bounds on their magnitudes are given~\cite{Fuentes-Martin:2020lea}. } \label{table:const}
    \end{table}}

We also follow the calculation of reference~\cite{Arnan:2019olv} to consider any sizeable contribution to the invisible width of the Z-boson-- particularly for $R_2$, where neutrino LQ couplings are not purely CKM generated. We enforce $2\sigma$ agreement for the observable effective number of neutrinos, $N_\nu$, in which contributions to this width are manifest (Table~\ref{table:const}).

Although electroweak constraints provide significant bounds on the top-muon coupling in these models, they are not a sensitive probe for the charm-electron coupling necessary to generate $\Delta a_e$. Loop-corrections to $Z\to \ell \ell$ from charm-containing loops scale proportional to the Z-boson threshold, $\propto m_Z^2/m_\phi^2$, notably without a top-mass enhancement. For constraining these couplings, we turn to complementary constraints from both high- and low-energy physics.

\subsubsection{\texorpdfstring{Contact Interactions and high $p_T$ leptonic tails}{Contact Interactions and high pT leptonic tails}}
\label{sec:ModelsB2}

Our model generates NP effects in dimension-6 SMEFT operators, contributing to $pp\to \ell \ell$ (dilepton production) and $pp\to \ell \nu$ (monolepton production), via tree-level t-channel processes. These can be probed directly in the high-$p_T$ tails of Drell-Yan processes at the LHC~\cite{Aaboud:2017buh}. For these processes, numerical analyses find minimal interference between NP effective operator contributions and \emph{a priori} several NP operators can be simultaneously constrained~\cite{Faroughy:2016osc, Greljo:2017vvb}. 

Under the assumption of no new degrees of freedom at or below the electroweak scale, NP contributions can be parameterised using the SMEFT: 
\begin{align}
    \mathcal{L}_\text{SMEFT} \supset \sum_\alpha C_\alpha \mathcal{O}_\alpha.
\end{align}
The analytic expressions for the relevant operators and corresponding WCs at high-scale ($m_\phi$) can be found in Table~\ref{table:SMEFT}. We emphasise that here we restrict our discussion to constraints on tree-level contributions to canonically dimension-6 SMEFT operators\footnote{As our models do not induce corrections to the W-vertex at tree-level, those constraints from reference~\cite{Fuentes-Martin:2020lea} are omitted from this discussion.}. As noted in Section~\ref{sec:ModelsB1}, these processes will be most relevant for constraining input `charm-electron' Yukawa couplings.

For second-generation quark processes generated in these models, effective interactions are also constrained by low-energy (semi)leptonic meson decays (Section~\ref{sec:ModelsB3}). Amplitude contributions to these high- and low-energy processes are related by a crossing-symmetry, and therefore provide complementary constraints\footnote{For a review of these limits in the context of charm decays and high-$p_T$ limits, see reference~\cite{Fuentes-Martin:2020lea}.}.

Note that with increased LHC luminosity, the allowed regions for the effective interactions manifest in high-$p_T$ leptonic tails are forecast to shrink significantly~\cite{Fuentes-Martin:2020lea, Greljo:2017vvb}, further restricting parameter space for both models. Here we limit our discussion to the present constraints.

\subsubsection*{Neutral currents and dilepton searches}

These processes provide the strongest constraints on the charged-current $cc\to ee$ transition generated at tree-level, constraining the magnitudes of Yukawa couplings responsible for correcting $\Delta a_e$. The WCs and associated operators constrained by these processes are indicated in Table~\ref{table:SMEFT}. For these processes, constraints on chirality-flipping scalar and tensor operators are irrelevant at this order, due to suppression from a light-fermion mass insertion. In Table~\ref{table:const} we quote the numerical upper bounds on explicit Yukawa couplings, derived from the two-sigma limits on the corresponding SMEFT WCs, using ATLAS 36.1 $\text{fb}^{-1}$ dataset~\cite{Aaboud:2017buh} -- Table 1 of reference~\cite{Greljo:2017vvb}. For correlated WCs, e.g. $C_{lq(1)}$ and $C_{lq(3)}$,  we take the most constraining of the two as a conservative bound.

Furthermore, in the $R_2$ model, tree-level contributions to the process $bb\to \mu\mu$ are generated via CKM mixing. Taking $V_{tb}\sim 1$, this process constrains the magnitude of the Yukawa coupling $y_{23}^{ReQ}$, resulting in the upper-bound quoted in Table~\ref{table:const}.

\begin{table*}[t]
\resizebox{12cm}{!}{%
  \def\arraystretch{1.8} 
    \begin{tabular}{|c|c|c|c|} 
    \hline
      \textsc{LQ, $\phi$}  & \textsc{SMEFT Operator}&\textsc{Wilson Coefficient} & \textsc{LHC high$-p_T$ constraints} \\
     \hline \hline
       $S_1$ &   $\mathcal{O}_{lq}^{(1)}=[\overline{L_{L}} \gamma_\mu L_{L}][\overline{Q_{L}}\gamma^{\mu} Q_{L}]$ &$C_{lq(1)}^{ijkl}=\frac{1}{4m_\phi^2} y^{SLQ,*}_{ik} y^{SLQ}_{jl}$ & $pp\to\ell^+ \ell^-$ \\ 
        &   $\mathcal{O}_{lq}^{(3)}=[\overline{L_L} \gamma_\mu \sigma_a L_L][\overline{Q_L}\gamma^{\mu} \sigma_a Q_L]$ &$C_{lq (3)}^{ijkl}=-C_{lq(1)}^{ijkl}$& $c\to d_k \overline{e}_i \nu_j$, $pp\to\ell^+ \ell^-$\\
         &  $\mathcal{O}_{eu}=[\overline{e_R} \gamma_\mu e_R][\overline{u_R}\gamma^{\mu} u_R]$  & $C_{eu}^{ijkl}=\frac{1}{2m_\phi^2} y^{Seu,*}_{ik} y^{Seu}_{jl}$& $pp\to\ell^+ \ell^-$\\
          & $\mathcal{O}_{lequ}^{(1)}=[\overline{L_L}e_R]i\sigma_2 [\overline{Q_L}u_R]^T$  & $C_{lequ(1)}^{ijkl}=\frac{1}{2m_\phi^2} y^{SLQ}_{ik} y^{Seu,*}_{jl} $ & $c\to d_k \overline{e}_i \nu_j$\\
           &   $\mathcal{O}_{lequ}^{(3)}=[\overline{L_L}\sigma_{\mu\nu}e_R]i\sigma_2 [\overline{Q_L}\sigma^{\mu\nu}u_R]^T $ &$C_{lequ(3)}^{ijkl}=-\frac{1}{4}C_{lequ(1)}^{ijkl}$ & $c\to d_k \overline{e}_i \nu_j$ \\
        \hline
      $R_2$ &   $\mathcal{O}_{lu}=[\overline{L_L} \gamma_\mu L_L][\overline{u_R}\gamma^{\mu} u_R]$ & $C_{lu}^{ijkl}=- \frac{1}{2 m_\phi^2} y^{RLu,*}_{jk}y^{RLu}_{il}$ &$pp\to\ell^+ \ell^-$\\
         &   $\mathcal{O}_{lequ}^{(1)}=[\overline{L_L}e_R]i\sigma_2 [\overline{Q_L}u_R]^T$ &$C_{lequ(1)}^{ijkl}= \frac{1}{2 m_\phi^2} y^{ReQ,*}_{jk}y^{RLu}_{il}$ &$c\to d_k \overline{e}_i \nu_j$\\
          &   $\mathcal{O}_{lequ}^{(3)}=[\overline{L_L}\sigma_{\mu\nu}e_R]i\sigma_2 [\overline{Q_L}\sigma^{\mu\nu}u_R]^T $& $C_{lequ(3)}^{ijkl}= \frac{1}{4} C_{lequ(1)}^{ijkl}$&$c\to d_k \overline{e}_i \nu_j$\\
           &   $\mathcal{O}_{qe}=[\overline{Q_L} \gamma_\mu Q_L][\overline{e_R}\gamma^{\mu} e_R]$ &$C_{qe}^{ijkl}=- \frac{1}{2 m_\phi^2} y^{ReQ,*}_{li}y^{ReQ}_{kj}$ &$pp\to\ell^+ \ell^-$\\
        \hline    
   \end{tabular}}
    \caption{Effective contact interactions in both LQ models. Labelling of WCs $(ijkl)$ indicates fermion flavour, and is ordered by appearance of the fermion in the effective operator. Following the SMEFT basis as outlined in reference \cite{deBlas:2017xtg}, $\sigma_a$ are the Pauli matrices ($a=1,2,3$) and $^T$ is understood to indicate transposition of the $SU(2)_L$ indices, exclusively. Note that we have adopted the SMEFT Warsaw-up basis, a variant of the standard Warsaw basis where the up-type quark mass matrix (rather than the down-type) is diagonal, and each of the non-physical rotation matrices appearing in \eqref{mappings} are set to the identity matrices. }
    \label{table:SMEFT}
    \end{table*}

\subsubsection*{Charged currents and monolepton searches}

Broadly speaking, high-$p_T$ monolepton searches give strong constraints on the relevant scalar and tensor operators. Constraints on the relevant WCs are given for rescaled versions of those quoted in Table~\ref{table:SMEFT}, defined via the tree-level matching conditions between the SMEFT and the low-energy effective theory~\cite{Fuentes-Martin:2020lea}:
\begin{equation}
\begin{aligned}
    \epsilon_{V_L}^{\alpha \beta i} &= - v^2 \frac{V_{ji}}{V_{2i}}[C_{lq(3)}^{\alpha \beta 2j}],\\
    \epsilon_{S_L}^{\alpha \beta i} &= - v^2 \frac{V_{ji}}{2V_{2i}}[C_{lequ(1)}^{\beta\alpha j2}]^*,\\
    \epsilon_{T}^{\alpha \beta i} &= - v^2 \frac{V_{ji}}{2V_{2i}}[C_{lequ(3)}^{\beta\alpha j2}]^*,
\end{aligned}\label{epsilonsSMEFT}
\end{equation}
where $v\sim 246$ GeV is the electroweak \emph{vev}, and $V_{ij}$ are CKM matrix elements. Note that $\epsilon_{V_L}$ is relevant for the $S_1$ model, but not for $R_2$. 

In Table~\ref{table:const}, we quote constraints on the relations in \eqref{epsilonsSMEFT}, derived via reference~\cite{Fuentes-Martin:2020lea} from the ATLAS 139 fb$^{-1}$\cite{Aad:2019wvl} and CMS 35.9 fb$^{-1}$\cite{Sirunyan:2018mpc} datasets. These limits will be numerically accounted for in our phenomenological study~(Section~\ref{sec:Pheno}).

\subsubsection{Mesonic decays: D-meson decays}
\label{sec:ModelsB3}

As the lepton Yukawa sector is explicitly flavour-decoupled in these models, this leads to mesonic constraints governed primarily by charged-current decays $D_{(s)}\to e \nu$. Contributions to these processes are generated by both models at tree-level.

\begin{figure*}[t!]
 \includegraphics[scale=1.6]{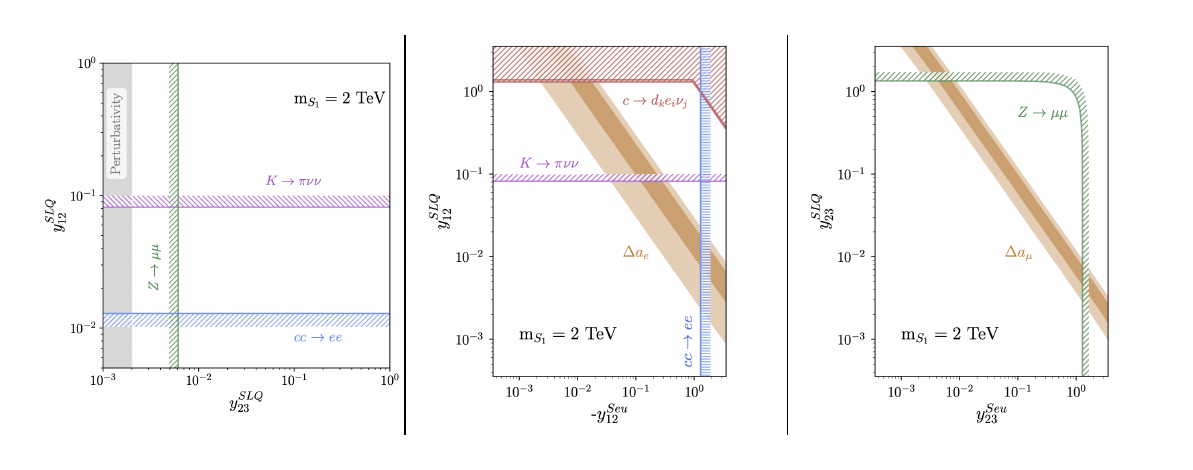}
\caption{$S_1$ model for fixed benchmark LQ mass of 2 TeV: hashed regions indicate that they are ruled-out by the labelled constraint, with conditions for each constraint outlined in Section III. The leftmost plot indicates the allowed region in the $y^{L}_{12}-y^{L}_{23}$ plane, with $y^R_{ij}$ fixed by the $\Delta a_{\ell}$ central values. Here, the shaded perturbativity constraint refers to that region being ruled-out by non-perturbative  generated right-handed couplings. The centre and rightmost plots show the relevant constraints for a scan over a decoupled muon- and electron-sector, and here the shaded region shows the one- and two-sigma allowed regions about the associated $\Delta a_\ell$ central values. 
}\label{fixedmassS1}
\end{figure*}

For both models in the considered parameter space, D-meson decay constraints are significantly weaker than the correlated high-$p_T$ leptonic constraints. We cross-checked Br($D_{(s)}\to e \nu$) constraints within two-sigma of their central values, for a sample of indicative allowed regions -- including, but not exclusive to, that shown in Figures~\ref{fixedmassS1} and~\ref{fixedmassR2}. We found that this subset of mesonic decays did not to rule out any additional parameter space, beyond what was already excluded. The experimental values used for these cross checks can be found in Table~\ref{table:const}.

\subsubsection{Mesonic decays: K-meson decays}
\label{sec:ModelsB4}
 Interactions involving Kaons are generated by these LQ couplings via $SU(2)_L$ invariance of the associated interaction, explicitly manifest in equations \eqref{lags} and \eqref{lagr}.
 
 Following the derivations in reference~\cite{Mandal:2019gff}, we re-derive the relevant constraints in terms of the Yukawas in the 'up-type' coupling regime. For this, we assume the Wolfenstein parameterisation for the CKM matrix. Namely the relevant combination of parameters for the decays considered here is:
 {\small
 \begin{equation}
     V_{cd}V_{cs} \sim - \lambda(1-\frac{1}{2} \lambda^2) \sim 0.22,
 \end{equation}}
 
\noindent where $\lambda\sim0.226$ is the standard Wolfenstein parameter\cite{PDG}. 
 
Contributions to the theoretically clean rare process Br$(K^+ \to \pi^+ \nu\overline{\nu})$ are found to give competitive constraints for the $S_1$ model. In Section~\ref{sec:ModelsA1} we highlighted the importance of this constraint for the case of a \emph{charmphilic} $(g-2)_\mu$ model, and it is similarly constraining for this coupling texture. As the effective interaction here is purely vector in nature, it does not run in QCD and has negligible electroweak running effects within the scope of this analysis. Following reference~\cite{Mandal:2019gff}, considering the dominant contribution at the $90\%$ confidence level gives the following upper bound:
\begin{align}
    |y^{SLQ}_{12}|\lesssim 4.09\times 10^{-2} \frac{m_\phi}{\text{TeV}}.\label{Knunu}
\end{align}
For real-valued Yukawa couplings, this process generates a more constraining bound for $S_1$ than those from $K-\overline{K}$ mixing or from the process $K_L\to \pi^0 \nu\overline{\nu}$~\cite{Kumar:2016omp}. For the $R_2$ model, contributions to this process are loop-order and the constraints are negligible relative to those otherwise considered.

Conversely, for the $R_2$ model the strongest constraint from the kaon sector come from tree-level leptoquark exchange to the helicity suppressed $K^0_L \to e^+ e^-$ transition.  Noting the vector nature of the effective contribution to this process, and following again from reference~\cite{Mandal:2019gff}, we derive an upper bound at $90\%$ confidence on the associated coupling:
\begin{align}
    |y^{ReQ}_{12}|\lesssim 9.5\times 10^{-2} \frac{m_\phi}{\text{TeV}}.\label{Kee}
\end{align}

We emphasise that for the purpose of these derivations we have assumed real-valued Yukawa couplings for each model, and recognise that complementary constraints may exist on any imaginary component if this assumption were to be relaxed. The $S_1$ leptoquark cannot contribute to this process at tree-level and higher order contributions  were found to be weaker than those from the high-$p_T$ LHC observables discussed in earlier subsections.

\section{\label{sec:Pheno} Phenomenology}
 \noindent In both models, the free parameters are summarised by \begin{equation}
\{y^{L}_{12}, y^{L}_{23}, y^{R}_{12},  y^{R}_{23} , m_{\phi}\},
\end{equation}
where we stress that the right- and left-handed couplings are explicitly defined for each model as per \eqref{coup1} and \eqref{coup2}. Presently the scalar leptoquark mass, $m_\phi$, is most strongly directly constrained using LHC searches for the decay of pairs of scalar LQ with couplings predominantly to first-generation leptons~\cite{Aad:2020iuy}: \begin{equation}m_{\phi}> 1.8\;\text{TeV}\; \text{at}\; 95\% \text{CL},\label{masslim}\end{equation} 
 providing a conservative lower bound\footnote{A novel study by the CMS collaboration considered searches for LQ with off-diagonal couplings to $t-\mu$~\cite{Sirunyan:2018ruf}, although this constraint is slightly weaker than that quoted above.}.

For the majority of this section, we will aim to fix the values of $\Delta a_\ell$ to a point within one-sigma of the central values, reducing the parameter degrees of freedom to $5-2=3.$ To achieve this, we scan logarithmically over positive perturbative left-handed couplings, and fix the right-handed couplings according to the full calculation of $\Delta a_\ell$ \eqref{deltaa}; the relative sign is absorbed into the allocation of right-handed couplings. Since this equation is a polynomial in $y^R_\ell$, and we restrict the couplings to be real-valued, we solve for the value of $y^R_\ell$ for which (\emph{a.}) Re($y^R_\ell$) is perturbative, and (\emph{b.}) Im($y^R_\ell$) is minimal. We truncate the input value for $y_{\ell}^R$ to be the real-component of this root. This is done algorithmically such that the input value is the closest fit to these requirements, listed in order of preference.  We then check the calculated value for $\Delta a_\ell$ and identify points which remain within one-sigma of the central values. 
 \begin{figure*}[t!]
\includegraphics[scale=1.6]{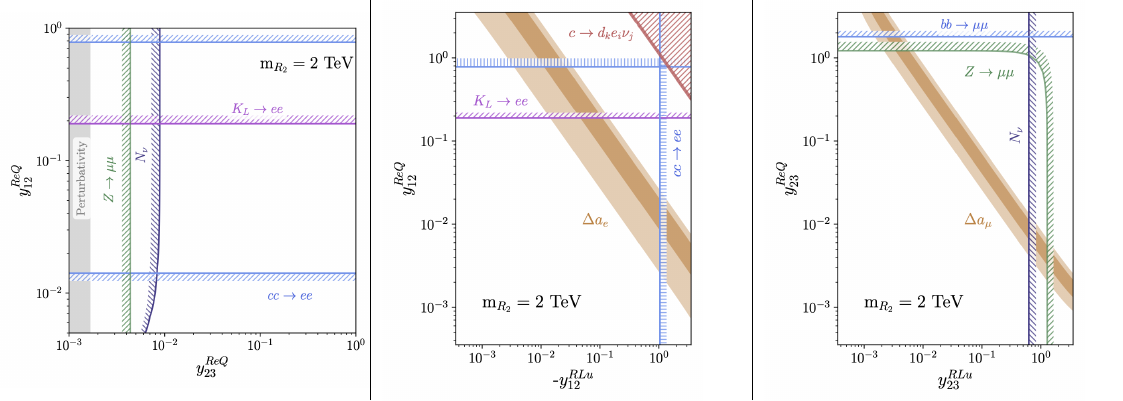}
\caption{$R_2$ model for fixed benchmark LQ mass of 2 TeV: hashed regions indicate that they are ruled-out by the labelled constraint, with conditions for each constraint outlined in Section III. The leftmost plot illustrates a scan over the $y^{L}_{12}-y^{L}_{23}$ plane, with $y^R_{ij}$ fixed by the $\Delta a_{\ell}$ central values. Here, the shaded perturbativity constraint refers to that region being ruled-out by non-perturbative  generated right-handed couplings. The centre and rightmost plots show the relevant constraints for a scan over a decoupled muon- and electron-sector, and here the shaded region shows the one- and two-sigma allowed regions about the associated $\Delta a_\ell$ central values.  
}\label{fixedmassR2}
\end{figure*}

The conditions for this parameter study are as follows, where experimental values implemented are as given in Table~\ref{table:const}. The below represent the requirements for values to pass constraints for each observable. 
\begin{enumerate}
\item $\Delta a_e$ and $\Delta a_\mu$ within 1 $\sigma$;
    \item Yukawa coupling perturbativity bound $|y|\leq \sqrt{4\pi}$;
    \item Effective $N_\nu$ within 2 $\sigma$;
    \item Effective left- and right-handed Z-muon couplings within 2$\sigma$ of their central values;
    \item LHC high-$p_T$ leptonic constraints, as listed in Table~\ref{table:const},
    \item and $m_{\phi}> 1.8\;\text{TeV}$.
\end{enumerate}
Furthermore, we impose the following model-specific constraints to the their respective parameter space:
\begin{itemize}
    \item$K^+\to\pi^+\nu\overline{\nu}$ derived coupling bound in equation~\eqref{Knunu} for $S_1$,
    \item and $K^0_L\to e^+ e^-$ bound in equation~\eqref{Kee} for $R_2$. 
\end{itemize} 

We will begin by identifying a benchmark region of parameter space for each model, then proceed to a full parameter grid-scan. We aim to achieve an upper-bound on the LQ masses capable of tackling the $(g-2)_{e/\mu}$ puzzle, and explore the emergent coupling structures of these solutions.

\subsection{\texorpdfstring{$S_1$ and $R_2$ models with $m_\phi=2$ TeV}
{S1 and R2 models with mass 2 TeV}}

We first begin to explore the available parameter space by performing grid-scans over coupling space projections, with a fixed $m_\phi=2$ TeV in both LQ models. The results of these scans are shown in in Figures ~\ref{fixedmassS1} and~\ref{fixedmassR2}. Constraints that do not explicitly appear in these figures were not competitive with those shown, within the illustrated region of interest.

\subsubsection{\texorpdfstring{Fixed about the $\Delta a_\ell$ central values}
{Fixed about the AMM central values}}

The leftmost plots illustrate the results of a scan fixing the right-handed couplings via the method outlined at the beginning of Section~\ref{sec:Pheno}. Starting with $S_1$, note particularly that contributions to effective $N_{\nu}$, via $Z\to~\nu\nu$, are suppressed significantly enough (\emph{i.e.} by CKM) to render these negligible for the considered parameter space. This is not true for $R_2$, and an effective $N_{\nu}$ bound is found to be competitive with that from $Z\to \mu\mu$. For the both LQs, this indicates a preferred allocation of couplings such that:
\begin{align}
    y^L_{12}\sim \mathcal{O}(10^{-2}), \;\text{and}\; 
    y^L_{23}\gtrsim \mathcal{O}(10^{-2}).
\end{align}
All points shown pass the requirement that $\Delta a_{e(\mu)}$ fall within the one-sigma region. These results act to demonstrate explicit existence of viable parameter space.

\subsubsection{\texorpdfstring{Decoupled $\mu$ and $e$ sectors}
{Decoupled muon and electron sectors}}
The centre and leftmost plots in Figures ~\ref{fixedmassS1} and~\ref{fixedmassR2} illustrate the planes of electron and muon couplings for each model. In each of these plots, the relevant $(y^L, y^R)_\ell$ couplings are scanned over perturbative values, without enforcing a fit with $\Delta a_\ell$. For these scans, the effectively decoupled parameters $(y^L, y^R)_{\ell'},$ where $\ell'\neq \ell$, are fixed to zero. Shaded regions illustrate the one- and two-sigma agreement with the $\Delta a_\ell$ values in \eqref{g-2mu} and \eqref{g-2e}. Hashed regions illustrate the strength of the important constraints on these parameter sub-space projections. These results demonstrate the impact of uncertainties on the $\Delta a_\ell$ measurements on these models' parameter space regions.

\subsection{\texorpdfstring{Parameter scan for $S_1$ and $R_2$ models}
{Parameter scan for S1 and R2 models}}

\begin{figure*}[t!]
\includegraphics[scale=0.9]{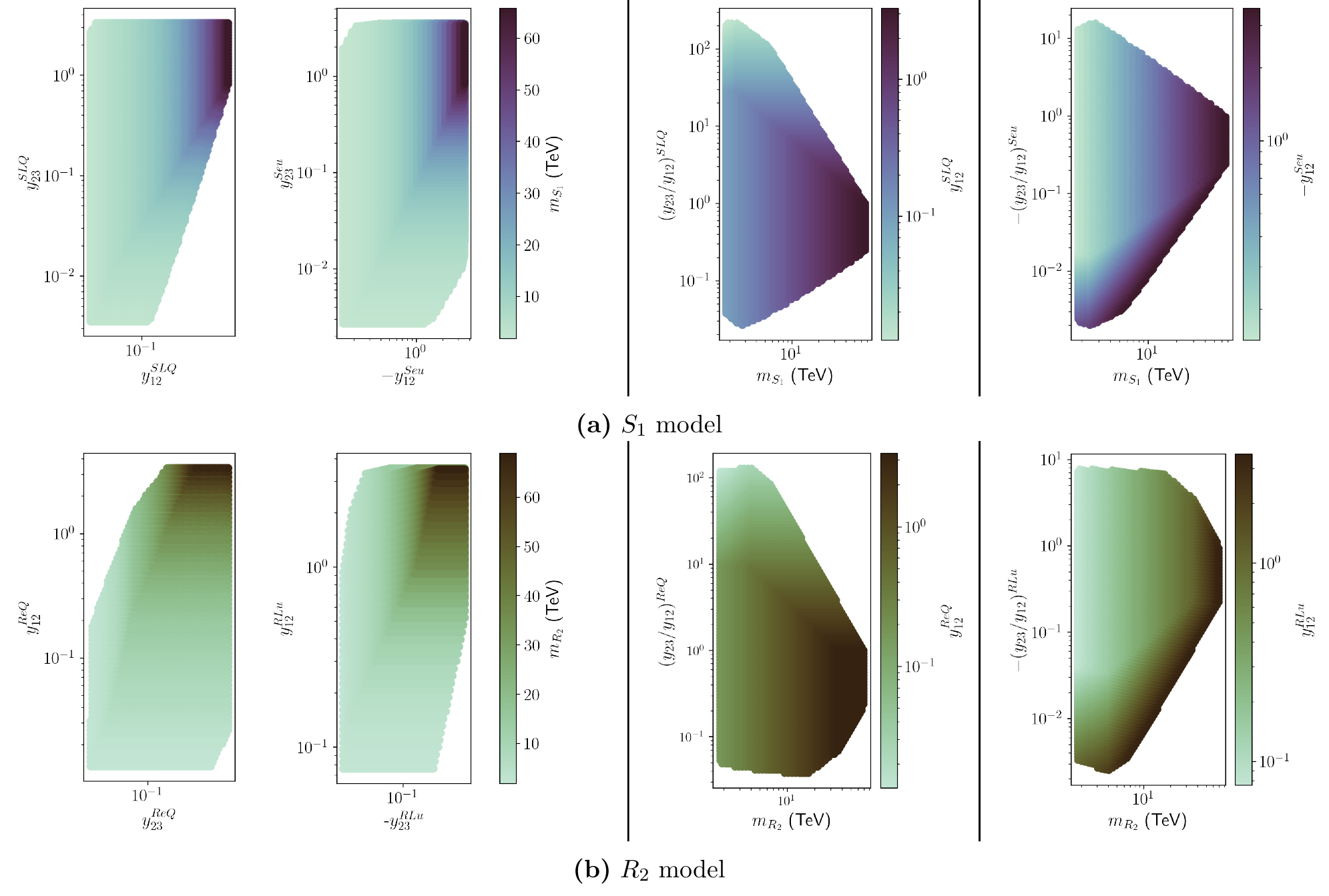}
\caption{Preferred coupling substructures for $S_1$ (a) and $R_2$ (b) $\Delta a_{e/\mu}$ model: these figures illustrate the results a grid scan of the coupling values for parameter points, performed as outlined at the beginning of Section III. Coloured points pass all constraints. In each subfigure, the left two plots demonstrate the preferred right- and left-handed coupling assignments for varying LQ mass, and share an associated colour bar. The right two plots show the ratio of nonzero couplings for both coupling matrices, as they vary with the LQ mass.}\label{S1R2tex}
\end{figure*}

Here we extend the benchmark region to include variation in the LQ mass for each model.
We logarithmically vary the parameters within the following ranges: 
\begin{equation}
   m_\phi \in [1.8, 100]\ \text{TeV}, \;\;\;\; y_L^{12}, y_L^{23} \in [0, \sqrt{4\pi}]\subset \mathbb{R}^+.
\end{equation}
The right-handed couplings are fixed, as per the algorithm for $\Delta a_\ell$ detailed in earlier discussion, and we require that points satisfy the constraints outlined at the beginning of Section III. We emphasise that the conclusions here are based on the current central values and error ranges for the anomalous magnetic moments, and should be revisited when new experimental results are obtained. The results of these scans are shown in Figure~\ref{S1R2tex}. We have labelled the couplings here explicitly in accordance with equations \eqref{lags} and \eqref{lagr}.

The left two plots in Figure \ref{S1R2tex} (a) and (b) can be thought of as illustrating the shrinkage and translation with $m_\phi$ of the constrained regions in the leftmost plots of Figures~\ref{fixedmassS1} and~\ref{fixedmassR2}. Here, the colourbar for these indicates the \emph{maximum} $m_\phi$ value that is allowed for that parameter combination; these masses are not unique for a particular coupling allocation because of the allowed region about the central values of $\Delta a_\ell$. As expected, Yukawa couplings show a trend towards higher magnitudes for larger LQ masses-- reflecting equations~\eqref{ALe} and \eqref{ALm}.  The upper limit on LQ mass for the $S_1$ and $R_2$ models under the specified constraints was found to be such that:
 \begin{align}
     m_{S_1}\lesssim 65\; \text{TeV},\;\text{and}\;
     m_{R_2}\lesssim 68\; \text{TeV}.
 \end{align}

In the right two plots in Figures \ref{S1R2tex} (a) and (b), the colourbars show the maximum allowed coupling value (labelled) for a particular mass and coupling ratio. For $S_1$, these narrow rapidly with increased mass $m_\phi$ to a preferred coupling ratio \begin{align} y_{23}/{y_{12}}\sim \mathcal{O}(10^{-1}),
 \end{align}
 for both right- and left-handed couplings. These plots for $R_2$ show a much less rapid convergence to a coupling ratio with increased mass $m_\phi$, but to the same approximate ratios. For smaller masses a wider range of coupling ratios are permitted. These ratios are consistent with the relations in equations~\eqref{bound1}-\eqref{bound4}, where we allow for the one-sigma regions about the central values of $\Delta a_{e/\mu}$.
 
 The rapid convergence of $S_1$ to this ratio would indicate that this LQ model structure is more sensitive to variations about the central values of $\Delta a_\ell$, and therefore improvements of these measurements (\emph{i.e.} those expected in the near future from the Muon $g-2$ collaboration~\cite{Chapelain:2017syu}) may necessitate a reassessment of the viability of such a model. This information is not immediately evident from contrast with centre and leftmost plots in Figures ~\ref{fixedmassS1} and~\ref{fixedmassR2}, however we note that the decoupled parameter planes do not reflect the influence of interplay between electron and muon couplings for relevant constraints.

Note that patterns emergent in the couplings within these datasets may motivate an algebraic relationship between LQ couplings of the quarks to charged-lepton generations. Known theoretical approaches can be explored to generate such hierarchical coupling structures in UV-complete NP models; for example, Froggatt-Nielson mechanisms~\cite{FROGGATT1979277}. We refrain from discussing this concept further here, but rather identify it as an avenue for future exploration.

 \subsection{\label{subsec:1.4}\texorpdfstring{Future of predicting $(g-2)_\tau$}{Future of predicting (g-2)tau}}

There are proposed measurement techniques for the quantity $(g-2)_\tau$~\cite{taummrev}, however up to the present there are only rough bounds on its experimental value~\cite{PDG};
\begin{equation}
   -0.052 <a_\tau<0.013\;\; (95\% \text{C.L}).
\end{equation}
The short lifetime of the tau makes measurement of this quantity notoriously difficult. By observation of the other two lepton flavours, we would expect that a measurement of this observable could provide further evidence for lepton non-universality in NP models. Furthermore, exploring the generation of hierarchical coupling structures could be further motivated by a fitting also to $(g-2)_\tau$, when such a measurement becomes available.

\section{Conclusions}

 We have argued that the apparent lepton-flavour universality violation in measurements of the anomalous magnetic moments of the electron \emph{and} muon can be explained using either of the mixed-chiral subset of scalar leptoquark models -- provided that Yukawa couplings are real-valued, and generated via an `up-type' mass basis. This mixed-chiral nature is necessary for the sign dependence of one-loop BSM corrections, consistent with the observed anomalies, and the `up-type' flavour ansatz avoids the explicit one-loop contributions to LFV processes such as $\mu\to e \gamma$. 
 
 There are precisely two scalar LQs which have the required coupling structure: the $SU(2)_L$ singlet $S_1\sim(\mathbf{3},\mathbf{1},-1/3)$ and the doublet $R_2\sim(\mathbf{3},\mathbf{2},7/6)$. We demonstrated the allowed parameter space, assuming real-values for Yukawa couplings, showing that both of these models are capable of generating $\Delta a_\mu$ and $\Delta a_e$ within $1\sigma$ of the current observed values, whilst also satisfying constraints from precision electroweak observables, LHC constraints and decays in the kaon sector --- illustrated explicitly for a benchmark mass of $m_\phi=2$ TeV in Figures~\ref{fixedmassS1} and~\ref{fixedmassR2}. A conservative upper bound on LQ mass in both models results in the following range of allowed masses: 
 \begin{align}
     1.8\lesssim \frac{m_\phi}{\text{TeV}} \lesssim 65.
 \end{align}

Hierarchical Yukawa coupling relationships may emerge in the allowed points for $S_1$ and $R_2$, depending on the particular choice of $m_\phi$. This motivates further study into possible texture generation mechanisms and their extension to models including effects in $(g-2)_\tau$.

The framework outlined in this paper can be easily adapted when new experimental results for $(g-2)_\ell$ are obtained, as expected in the near-future from experiments such as Muon $g-2$ \cite{Chapelain:2017syu}. We motivate the ongoing consideration of mixed-chiral LQs for flavour phenomenology in new physics models, particularly those permitting violation of lepton-flavor universality.

\appendix

{\small 
\subsection*{\texorpdfstring{\bf A note on the $\Delta a_\mu$ theoretical prediction}{Note on gm2 theoretical calculation}}
After the initial version of work was completed, a result from the BMWc collaboration~\cite{Borsanyi:2020mff} was released giving an improvement on the theoretical prediction for hadronic vacuum polarization contribution to $(g-2)_\mu$ from lattice QCD. Their result indicates that within the current experimental measurements, NP explanations in this quantity may be entirely unnecessary. In the $\Delta a_\mu$ result quoted in equation~\eqref{g-2mu}, the hadronic contribution to the corresponding theoretical calculation originates from the averaged $R-$ratio procedural results for $(g-2)_\mu$~\cite{Jegerlehner:2017lbd,Keshavarzi:2019abf,Davier:2019can}. Prior to this BMWc result, the $R-$ratio procedure was highly favoured over lattice methods -- due to lattice methods having very large associated theoretical uncertainties. Reference~\cite{Borsanyi:2020mff} is the first to achieve competitive precision for the lattice method; however, as the authors themselves acknowledge, this result is yet to be independently verified. For this reason, we use the $R-$ratio theoretical values for our study.}

\vspace{0.2cm}
\begin{acknowledgments}
 This work was supported in part by the Australian Research Council and the Australian Government Research Training Program Scholarship initiative. IB would like to acknowledge Rupert Coy, Peter Cox, John Gargalionis and Jonathan Kriewald for useful discussion, and the Gwenneth Nancy Head Foundation for funding attendance at the SLAC Summer Institute which helped to inspire this work.
\end{acknowledgments}

\bibliographystyle{aapmrev4-2.bst}
\bibliography{manuscript}

\end{document}